%% file: acl_latex.tex
\pdfoutput=1

\documentclass[11pt]{article}

\usepackage[]{acl}

\usepackage{times}
\usepackage{latexsym}

\usepackage{graphicx}
\usepackage{todonotes}

\usepackage[T1]{fontenc}

\usepackage[utf8]{inputenc}

\usepackage{microtype}

\usepackage{inconsolata}

\usepackage[normalem]{ulem}
\usepackage{subfig}
\usepackage{adjustbox}
\usepackage{multicol}
\usepackage{multirow}
\usepackage{booktabs}
\usepackage{hyperref}
\usepackage{listings}
\usepackage{enumitem}
\usepackage{comment}
\usepackage{wasysym}
\usepackage{pifont}
\hypersetup{
    colorlinks=true,
    linkcolor=blue,
    filecolor=magenta,      
    urlcolor=cyan,
    }
\usepackage{dialogue}
\usepackage[framemethod=tikz]{mdframed}

\title{Clarifying the Path to User Satisfaction: \\ An Investigation into Clarification Usefulness}
\author{Hossein A.~Rahmani\textsuperscript{\ding{170}}\thanks{\ \ Corresponding author}~ Xi Wang\textsuperscript{\ding{170}} Mohammad Aliannejadi\textsuperscript{\ding{171}} \\ \textbf{Mohammadmehdi Naghiaei}\textsuperscript{\ding{168}} \textbf{Emine Yilmaz}\textsuperscript{\ding{170}} \\ {\textsuperscript{\ding{170}}University College London, London, UK} \\ {\textsuperscript{\ding{171}}University of Amsterdam, Amsterdam, The Netherlands} \\ {\textsuperscript{\ding{168}}University of Southern California, California, USA} \\ \texttt{{\{hossein.rahmani.22,xi-wang,emine.yilmaz\}@ucl.ac.uk}} \\ \texttt{m.aliannejadi@uva.nl, naghiaei@usc.edu}}

\begin{document}
\maketitle
\begin{abstract}
Clarifying questions are an integral component of modern information retrieval systems, directly impacting user satisfaction and overall system performance. Poorly formulated questions can lead to user frustration and confusion, negatively affecting the system's performance. This research addresses the urgent need to identify and leverage key features that contribute to the classification of clarifying questions, enhancing user satisfaction. To gain deeper insights into how different features influence user satisfaction, we conduct a comprehensive analysis, considering a broad spectrum of lexical, semantic, and statistical features, such as question length and sentiment polarity. Our empirical results provide three main insights into the qualities of effective query clarification: (1) specific questions are more effective than generic ones; (2) the subjectivity and emotional tone of a question play a role; and (3) shorter and more ambiguous queries benefit significantly from clarification. Based on these insights, we implement feature-integrated user satisfaction prediction using various classifiers, both traditional and neural-based, including random forest, BERT, and large language models. Our experiments show a consistent and significant improvement, particularly in traditional classifiers, with a minimum performance boost of 45\%. This study presents invaluable guidelines for refining the formulation of clarifying questions and enhancing both user satisfaction and system performance.
\end{abstract}

\input{sections/01-introduction.tex}
\input{sections/02-related-work}

\input{sections/02-experimental-setup.tex}
\input{sections/03-analysis.tex}
\input{sections/04-results.tex}
\input{sections/05-conclusion}

\section*{Limitations}
In this paper, we delve into the significance of query and clarifying question features within a clarifying question system, aiming to enhance the utility of these questions and ultimately elevate user satisfaction. Nonetheless, our research faces constraints from the restricted publicly available resources, which requires more extensive datasets in future research studies. Moreover, the availability of resources also resulted in our conclusions exclusively to the Bing search platform, although we have taken steps to mitigate this limitation through our conducted user study.

\section*{Acknowledgements}
This research is supported by the Engineering and Physical Sciences Research Council [EP/S021566/1], the Alan Turing Institute under the EPSRC grant [EP/N510129/1] and the EPSRC Fellowship titled ``Task Based Information Retrieval'' [EP/P024289/1]. Any opinions, findings, conclusions or recommendations expressed in this material are those of the authors and do not necessarily reflect those of the sponsors.

\bibliography{anthology,custom}

\clearpage
\newpage

\appendix

\section{GPT-4 Prompts for CQ usefulness classification}
\label{sec:prompts}
\begin{mdframed}
\begin{dialogue}
    \speak{System} In a mixed-initiative conversational search system, a user's query might be ambiguous, and the system can ask a clarifying question to clarify the user's information need. In a real system, user satisfaction with the clarifying question is a very important task that should be considered. The prediction is a classification with three classes including: (1) Good, (2) Fair, and (3) Bad. In summary, this indicates that a Good clarifying question should accurately address and clarify different intents of the query. It should be fluent and grammatically correct. If a question fails in satisfying any of these factors but still is an acceptable clarifying question, it should be given a Fair label. Otherwise, a Bad label should be assigned to the question.
    \speak{query} Given the details about the satisfaction of a clarifying question, predict only the label for the following query, clarifying question, and the options for the clarification response: Query: `\{\}', clarifying question: `\{\}'.
\end{dialogue}
\end{mdframed}

\section{User Study Guidelines}
Here, we detail the \textit{instructions} that we present to the domain experts for another comprehensive evaluation of features that could contribute to the usefulness of clarifying questions:

\noindent\rule{7.5cm}{1pt}
\noindent\textbf{
    \begin{center}
        User Study Instructions
    \end{center}
}
\noindent\rule{7.5cm}{1pt}

This user study stands upon the research domain of asking clarifying questions, which aims to provide appropriate clarifying questions when an information-seeking system encounters ambiguous queries and needs to reveal users’ true intents. Therefore, in this user study, we aim to investigate the users’ opinions towards which features they value for the usefulness of a clarifying question. For example, a user could argue the necessity of a clarifying question is natural by itself and includes novel information compared to a given query.  

To collect the corresponding feedback from users, we ask you to take two stages of action. First, you need to label if a clarifying question is considered useful or not in general. To do so, you only check the checkbox if you consider a clarifying question useful. Next, you select features that contribute to a useful clarifying question or the ones that are missing and make the corresponding clarifying question unuseful. We prefer the selection of multiple features if they are considered valuable.

The considered features are categorised into two groups:

\begin{enumerate}
    \item Clarifying Question-only Features
    \begin{itemize}
        \item \textbf{Naturalness:} If a clarifying question is natural if it looks like a proper question in revealing the real intent given by the corresponding query.
        \item \textbf{Grammar:} The clarifying question is written in correct grammar.
        \item \textbf{Fluency:} The clarifying question is written in fluent English.
        \item \textbf{Question Template:} If the clarifying question is useful since it uses a particular question template or vice versa.
    \end{itemize}
    \item Features on Query and CQs
    \begin{itemize}
        \item \textbf{Coverage:} The clarifying question extends the query by covering the required aspects, which enables the system to identify relevant information.
        \item \textbf{Relevance:} The clarifying question is related to the corresponding query.
        \item \textbf{Novelty:} The clarifying question identifies the new aspects that are not mentioned in the query. Different from the coverage, for novelty, we value the necessity of including new aspects instead of a full consideration of related aspects.
        \item \textbf{Efficiency:} The ability of a clarifying question can save time for exploration and help in identifying the relevant information.
    \end{itemize}
\end{enumerate}

\end{document}

%% file: sections/01-introduction.tex
\section{Introduction}
\label{sec:introduction}
Asking clarifying questions (CQs) plays a pivotal role in enhancing both conversational search~\cite{AliannejadiSigir19} and web search experiences~\cite{zamani2020generating}. Timely and high-quality questions can significantly improve system performance~\cite{krasakis2020analysing} as well as overall user experience~\cite{kiesel2018toward,shi2022learning}. However, the adverse effects of poorly timed~\cite{aliannejadi2021building} or inappropriate questions can be significant, often leading to user frustration and dissatisfaction~\cite{zou2023users}. Given these challenges, optimizing the formulation of CQs has become an area of growing research interest.

Much research has studied the effectiveness of CQs in improved retrieval performance~\cite{krasakis2020analysing,aliannejadi2021analysing,cast2022,aliannejadi2020convai3,hashemi2020guided,shi2023and}. For example, \cite{krasakis2020analysing} studies different types of CQs and their answers, such as positive or negative answers, to characterize their impact on retrieval performance. TREC CAsT, in its latest edition in 2022~\cite{cast2022}, includes mixed-initiative conversation trajectories and features an independent mixed-initiative subtask, mainly focusing on search clarification. Several models are proposed in the ConvAI3 challenge~\cite{aliannejadi2020convai3}, aiming to incorporate CQs in the ranking process, mostly proposed based on pre-trained language models. Complementing this focus, some research integrates ranking and clarification features within learning objectives~\cite{hashemi2020guided}, while others explore the inherent risks by gauging the prospective retrieval gains~\cite{wang2021controlling}.
In the information retrieval (IR) community, there is a long-standing discussion suggesting that superior system performance in terms of relevance does not necessarily result in enhanced user experience or usefulness~\cite{mao2016when}. This has catalyzed a distinct line of research focused on comprehending the user experience with CQs~\cite{kiesel2018toward,zou2023users,zou2023asking,siro2022understanding,zamani2020analyzing,tavakoli2022mimics}.

It is pertinent to note that, in this study, we categorize ``useful clarifying questions'' as those that lead to higher user satisfaction.
Specifically, we argue that users' overall satisfaction depends on a variety of facets of a triad: the query, its CQs, and the corresponding candidate answers. This perspective is motivated by a recent study~\cite{siro2022understanding} that focuses on user satisfaction in task-oriented dialogues, emphasizing the importance of utterance relevance and efficiency. While there is existing research, such as that by \citet{tavakoli2022mimics} and \citet{zamani2020mimics}, that models user interaction and engagement with clarification panes, these studies primarily offer observational insights and have produced publicly available datasets like MIMICS and MIMICS-Duo. In contrast to these studies, our focus shifts toward predicting the practical value -- usefulness and user satisfaction -- of CQs, based on various attributes of search queries, CQs, and their candidate answers.

In summary, much of the existing research has concentrated on the quality and effectiveness of CQs in the context of retrieval gain. However, there is a noticeable gap in characterizing and predicting the real-world applicability or `usefulness' of these questions. The concept of usefulness is intricately connected to user satisfaction, as underscored by \citet{siro2022understanding}. Addressing this gap is challenging due to the multitude of factors influencing user experience beyond mere relevance~\cite{mao2016when}. To tackle this unexplored aspect of CQs, our study aims to answer the following research questions:

\begin{enumerate}[label=\textbf{RQ\arabic*},nosep,leftmargin=*]
    \item What features of clarifying questions help achieve higher user satisfaction? \label{RQ1}
    \item For which search queries do users prefer to use clarification? \label{RQ2}
    \item What is the impact of each feature on the usefulness prediction of clarifying questions? \label{RQ3}
\end{enumerate}

To this end, we conduct a comprehensive analysis and demonstrate their effectiveness in predicting question usefulness.\footnote{\url{https://github.com/rahmanidashti/CQSatisfaction}} In particular, we analyze the characteristics of CQs and user queries on two widely used real-world datasets, namely, MIMICS~\cite{zamani2020mimics} and MIMICS-Duo~\cite{tavakoli2022mimics}. The choice of using these two datasets is grounded on a recent survey~\cite{hossein2023survey}, indicating that MIMICS and MIMICS-Duo are the only two datasets allowing the evaluation of clarifying question usefulness as per user satisfaction levels. Leveraging these two datasets, we conduct a comprehensive evaluation over multiple dimensions, including the template structures of CQs, the number of candidate answers available, subjectivity and sentiment polarity of CQs, the length of both CQs and queries, query ambiguity, as well as the predicted relevance between CQs and queries. To augment the evaluation of useful CQs, we further conduct a user study over a number of features, such as question naturalness. In addition, to show the benefit of the learned relationships between numerous aforementioned features and CQ usefulness, we leverage the extracted features and feed them to multiple classifiers to predict CQ usefulness, leading to significant performance improvement.

Therefore, the main contributions of our work are as follows:
\begin{itemize}[nosep,leftmargin=*]
    \item A comprehensive exploration of relevant features that could contribute to the accurate classification of useful clarifying questions.
    \item Rich analysis of aspect-focused, long, sentimental positive, and subjective clarifying questions, demonstrating their positive effect on usefulness.
    \item Using positively correlated features to achieve significant improvements on both traditional and advanced machine learning classifiers, leading to large improvements (e.g., Precision of 0.9658)
\end{itemize}

%% file: sections/02-related-work.tex
\section{Related Work}
In this section, we discuss the existing research that pertains to the domain of asking clarifying questions (ACQ) in a conversational information-seeking system. Although there have been previous efforts in this area, none of them has specifically examined the potential features that contribute to the usefulness of clarifying questions. Therefore, we reviewed the related literature to provide a background description of our contributions in this paper.

Benefiting from the released public datasets with available query-clarifying question relevance labels, such as Qulac \cite{AliannejadiSigir19} and ClariQ~\cite{aliannejadi2021building}, many clarifying question ranking models have been introduced~\cite{kumar2020ranking,rao2019answer}. For example, in \cite{kumar2020ranking}, with concatenated embeddings of posts, clarifying questions as well as optional answers from a StackExchange-based dataset \cite{rao2018learning} as input to a multi-layer neural model, they estimate the probability of a clarifying question being relevant or not. 
However, due to the diverse and complex nature of clarifying questions, it is challenging to effectively address this asking clarifying question task in a retrieval manner \cite{zamani2020generating,zhao2022generating,sekulic2021towards}. In particular, to enable the generation of appropriate clarifying question, a good comprehension of the queries and their likely intents is required. For example, \citet{zamani2020generating} specifically designed a query aspect modelling module as well as multiple query aspect encoders to encompass the information within queries for clarification generation effectively. So far, the existing studies illustrate the effectiveness of their generated clarifying questions by comparing to the available ground-truth \cite{sekulic2021towards}, or human annotators \cite{zamani2020generating}. However, there is limited effort in exploring the aspects or features about a useful clarifying question. A similar contribution is \citet{siro2022understanding}, which evaluate the aspects of dialogues that could improve the user satisfaction level in a conversational recommendation scenario. Therefore, we argue that the investigation on revealing aspects for evaluating the usefulness of clarifying questions can guide the future development of clarifying question generation.

%% file: sections/02-experimental-setup.tex
\section{Experimental Setup}
\label{sec:experimental_setup}
In this study, we investigate numerous features that likely contribute to the usefulness of clarifying questions. In conversational information-seeking systems, users often submit diverse types of queries, ranging from statements to questions, varying in length (short or long). A clarifying question can be returned by the corresponding system to better reveal users' true information needs based on the query-as-input from the end users. Intuitively, to assess the usefulness of a clarifying question, we should not rely solely on the question itself. It is crucial to jointly model both the query and the corresponding clarifying question. Meanwhile, to examine user satisfaction with the presented clarifying questions, we leverage two commonly used datasets, MIMICS and MIMICS-Duo, which encompass the corresponding labels. Table~\ref{tab:dataset_stat} presents a statistical summary of these datasets. Moreover, with these two datasets, we assess the utility of various features, including query-oriented and clarifying question-independent features. These two datasets are the only real-world clarification datasets available, as highlighted in a recent survey on asking clarification questions datasets \cite{rahmani-etal-2023-survey}. These datasets are derived from Microsoft Bing, a widely recognised search engine, lending a degree of real-world applicability to our findings.

For the question-based features, we consider (1) the question template variance, (2) clarifying question presentation with a varied number of candidate answers, (3) question subjectivity, (4) sentimental polarity of questions and (5) question length. As for the query-oriented features, we investigate the impact of (6) query length in words, (7) query types (ambiguous or faceted) and (8) query-question relevance.
Note that partial features, such as the length of questions and the number of candidate answers, were studied in \cite{zamani2020mimics}. However, these features remain underexplored when it comes to providing comprehensive insights into the usefulness of CQs. Therefore, in this study, we extend the observations to the two datasets, systematically explore many other potential features and develop classifiers for the prediction as promising guidance for the future development of clarifying questions.
Note that, for the quantification of each feature, we detail the strategy in each of their corresponding discussions.

Specifically, while comparing the contributions of features, we observe a common issue of data imbalance -- the number of positive queries does not equal to negative ones. To address this issue, we normalize the scores of evaluated features based on the frequency of the corresponding groups. For instance, if 60 positive labels are assigned to 100 long clarifying questions and 15 positive labels are assigned to 50 short clarifying questions, we score the long and short questions 0.6 and 0.3, respectively, for comparison.

In the second part of this study, we investigate the value of the learned features from the previous step on classifying the usefulness of clarifying questions. To do so, we develop and explore numerous machine learning classifiers to estimate the usefulness of a given clarifying question. For evaluation, we partition each dataset into 80\% as the training set and the rest 20\% as the test set. The experimented approaches are from traditional machine learning and recent neural classifiers.

For the classic approaches, we consider Decision Tree Classifier (DTC) \cite{breiman2017classification}, Random Forest Classifier (RFC) \cite{breiman2001random} and Support Vector Classifier (SVC) \cite{fan2008liblinear} with a linear kernel. For neural approaches, we encode the input using pre-trained language models, including:
\begin{itemize}
    \item \textbf{BERT} \cite{jacob2019bert}, a transformer-based model which reads text bi-directionally, capturing deep contextual information from both directions. 
    \item \textbf{DistilBERT (DBT)} \cite{victor2019distilbert}, a lighter version of BERT via knowledge distillation with 40\% fewer parameters.
    \item \textbf{ALBERT} \cite{Zhenzhong2020albert}, another lighter version of BERT by employing factorised embedding parameterization and cross-layer parameter sharing, trained with an additional inter-sentence coherence loss to the masked language modelling loss that was used for training BERT.
    \item \textbf{BART} \cite{lewis2020bart}, it combines auto-regressive and auto-encoding training, pre-training by corrupting and then reconstructing sentences.
    \item \textbf{GPT-4}, the latest variant of the GPT-series models \cite{radford2018improving}, which has shown its advance in various language modelling tasks. We deploy a prompt learning method for classifying the usefulness of clarifying questions. The corresponding prompt is detailed in the appendix \ref{sec:prompts}.
\end{itemize}

Traditional machine learning models take in TF-IDF weighted bag-of-word features as input, which are extracted from the text data. 
We implemented these models using popular libraries such as Scikit-learn\footnote{\url{https://scikit-learn.org/stable/}} \cite{scikit-learn}, HuggingFace\footnote{\url{https://huggingface.co/}} \cite{wolf2020transformers}, and PyTorch\footnote{\url{https://pytorch.org/}} \cite{paszke2019pytorch}. To assess the performance of our models, we used standard evaluation metrics for supervised classification tasks, including Precision, Recall, and F1 score. All of the implementations, parameters, and datasets can be found on our GitHub repository.

%% file: sections/03-analysis.tex
\section{Clarification Usefulness}
\label{sec:analysis}
In this section, we aim to answer \ref{RQ1} and \ref{RQ2} at first by examining various potential factors and characteristics of CQs and queries that are pertinent to the effectiveness and usefulness of a clarifying question while applied to a query. The first part of the feature effectiveness examination focuses on the independent investigation of the clarifying questions themselves without taking the corresponding queries into account. The involved features include question template variants, number of candidate answers, subjectivity and sentiment polarity of questions. Next, we further examine the features of query differences as well as the relationships between query and clarifying questions in the second part.

\begin{table}
\centering
\caption{Dataset statistical summary. Question-label refers to the human-labeled usefulness of a clarifying question.}
\resizebox{\columnwidth}{!}{
\begin{tabular}{lll}
\toprule
  & \textbf{MIMICS-Manual} & \textbf{MIMICS-Duo} \\
  \midrule
\# unique queries & 2,464             & 306          \\
\# unique CQs & 252 & 22 \\
\# query-clarification pairs & 2,832             & 1,034          \\
\# question-label & 575             & 1,034         \\
\bottomrule
\label{tab:dataset_stat}
\end{tabular}}
\end{table}

\subsection{Characterizing Clarifications with Usefulness Rate}

\begin{table}
\centering
\caption{Clarifying questions templates on MIMICS and MIMICS-Duo with CQ quality labels. No bad label is given to the CQs with the following templates.}
\resizebox{\columnwidth}{!}{
\begin{tabular}{p{0.40\columnwidth}| p{0.12\columnwidth}|p{0.12\columnwidth}| p{0.12\columnwidth}|p{0.12\columnwidth}|p{0.12\columnwidth}|} 
\toprule
\multirow{2}{*}{\textbf{CQ Template}}  & \multicolumn{2}{c}{\textbf{MIMICS}} & \multicolumn{2}{c}{\textbf{MIMICS-Duo}} & \\
\cmidrule{2-6}
  & \textbf{Good} & \textbf{Fair} & \textbf{Good} & \textbf{Fair} & Comb. \\ 
\midrule
What (would you like | do you want) to do with \_\_\_\_? & 1.0 & 0.0 & 1.0  & 0.0 & 2.0 \\
\hline
What (would you like | do you want) to know about \_\_\_\_? & 0.9367 & 0.0632  & 0.75  & 0.25 & 1.6867\\
\hline
(Which|What) \_\_\_\_ are you looking for? & 0.6818 & 0.3181 & 0.8333 & 0.1666 & 1.5151\\
\hline
(Which|What) \_\_\_\_ do you mean? & 1.0  & 0.0 & 0.5 & 0.5 & 1.5  \\
\hline
What are you trying to do? & 0.0 & 1.0 & 1.0 & 0.0 & 1.0\\

\hline
Who are you shopping for? & 0.5714 & 0.4285 & -  & - & 0.5714 \\
\hline
Do you have \_\_\_\_ in mind? & 0.5 & 0.5 & -  & - & 0.5 \\

\bottomrule
\end{tabular}}
\label{tbl:cq_template}
\end{table}

\subsubsection{Question Templates}
\label{ssec:template}
A clarifying question can take various forms, yet convey the same meaning. Indeed, with an example query of ``monitor'', both ``(Which/What) [monitor] are you looking for'' and ``What (would you like | do you want) to know about [monitor]?'' can be used. Essentially, to reveal the true intent behind a user's query, there are diverse formats or templates that can be deployed to shape a clarifying question for optimised performance. In the literature, \citet{zamani2020generating} recently proposed to generate a majority of clarification types in a pre-existing set of question templates. In this study, to identify the most effective templates, we analyze both datasets and focus on those clarifying questions with top frequent formats. Table \ref{tbl:cq_template} presents the average usefulness of each template with respect to each label. We sort the templates in order of the sum of \textit{Good} scores in both datasets. Based on the table, question templates seeking detailed information consistently yield higher user satisfaction than those that simply rephrase user needs. For example, ``What would you like to know about [QUERY]?'', are found to be more useful than those that ask questions like ``What are you trying to do?'' or ``Who are you shopping for?''. 
A simple rephrasing request from a clarifying question could consume the user's patience in continuing the search and lower the level of user satisfaction. Instead, by having clarifying questions asking for specific facets of user intent, it enables the user to effectively augment the initial query with enriched information and improve the likelihood of retrieving relevant information. This finding aligns with the observations in the literature that users are more satisfied with those questions that they can foresee the benefit of answering them \cite{zou2023users}.

\subsubsection{Number of Candidate Answers}

\begin{table}
\centering
\caption{Satisfaction level for clarification panes per number of candidate answers (options).}
\resizebox{\columnwidth}{!}{
\begin{tabular}{lllllll}
\toprule
Dataset & Label & \#2 & \#3 & \#4 & \#5 \\
\midrule
\multirow{3}{*}{MIMICS} & Bad & 0.0 & 0.0 & 0.0 & 0.0\\
 & Fair & 0.1117 & 0.0538 & 0.0728 & 0.1509 \\
 & Good & 0.0517 & \textbf{0.1625} & 0.1236 & \textit{0.1361}\\ \midrule
\multirow{3}{*}{MIMICS-Duo} & Bad & 0.0485 & 0.0333 & 0.0225 & 0.0229 \\ 
 & Fair & 0.3059 & 0.2208 & 0.1412 & 0.1289 \\
 & Good & 0.6455 & 0.7458 & 0.8361 & \textbf{0.8481} \\
\bottomrule
\end{tabular}}
\label{tbl:candidate_answers_len}
\end{table}

To augment the presentation of a clarifying question, some search engine services, like Bing, also add a number of candidate answers to simplify the users' task in phrasing answers and improve users' experience. However, the optimal number of candidate answers to be presented remains underexplored. In Table \ref{tbl:candidate_answers_len}, we illustrate the range of candidate answers in the clarification pane, which varies from two to five in both MIMICS and MIMICS-Duo. The table also presents the clarification usefulness per label and the number of candidate answers. The results show that clarification panes with only two candidate answers receive low user satisfaction on both datasets, and a close satisfaction level can be observed with more candidate answers without a consistent optimal number (3 for MIMICS and 5 for MIMICS-Duo). In particular, the use of any 3 to 5 answers can consistently outperform the use of 2 answers. This indicates a requirement to involve rich aspects as an extension for the submitted query for users to interactively indicate their true intent. Users are more satisfied with diverse Clarifying questions, as the candidate answers in the clarification pane help provide more Clarifying questions.
One or two candidate answers do not sufficiently cover all the aspects of the query and user needs. Given the clarifying question representation manner of leveraging candidate answers, we show that there is a threshold of offering more than two candidate answers towards a positive user experience. This finding is consistent with \citet{zamani2020analyzing}, who also explore the relationship between candidate answers and user engagement.

\begin{figure}%
    \centering
    \subfloat[\centering Queries]{{\includegraphics[width=.45\columnwidth]{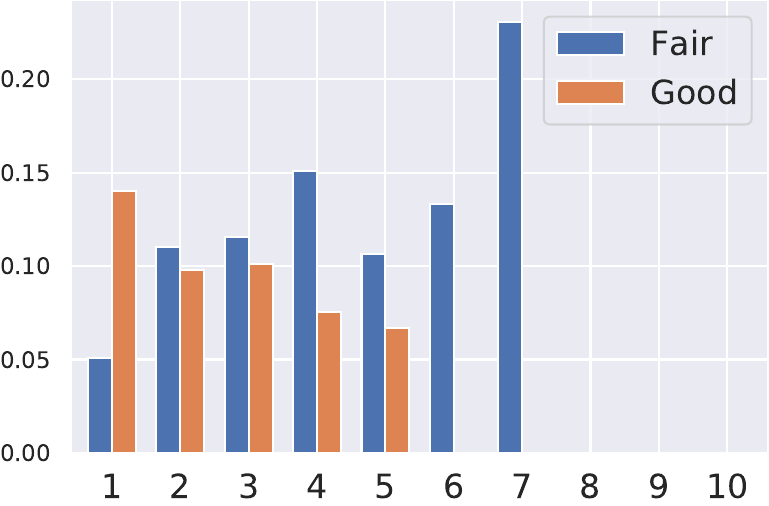} }}%
    \quad
    \subfloat[\centering Questions]{{\includegraphics[width=.45\columnwidth]{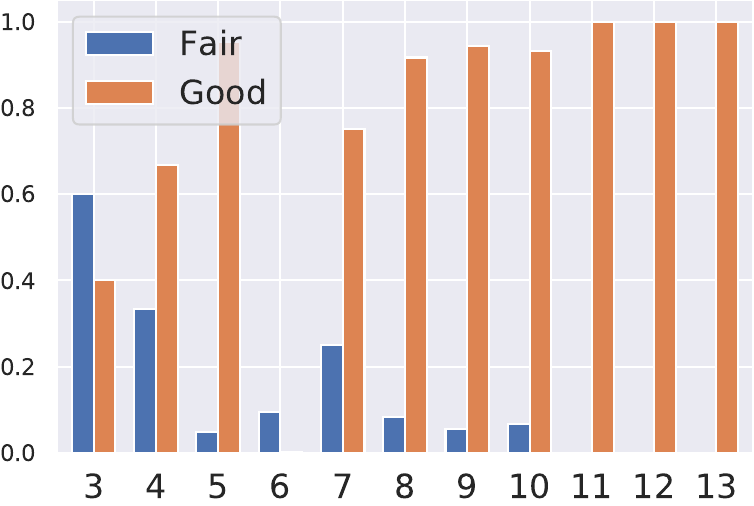} }}%
    \caption{clarifying question usefulness according to the length of queries and questions on MIMICS (similar pattern on MIMICS-Duo).}%
    \label{fig:mimics_length_to_label}%
\end{figure}

\begin{figure}%
    \centering
    \subfloat[\centering MIMICS]{{\includegraphics[width=.45\columnwidth]{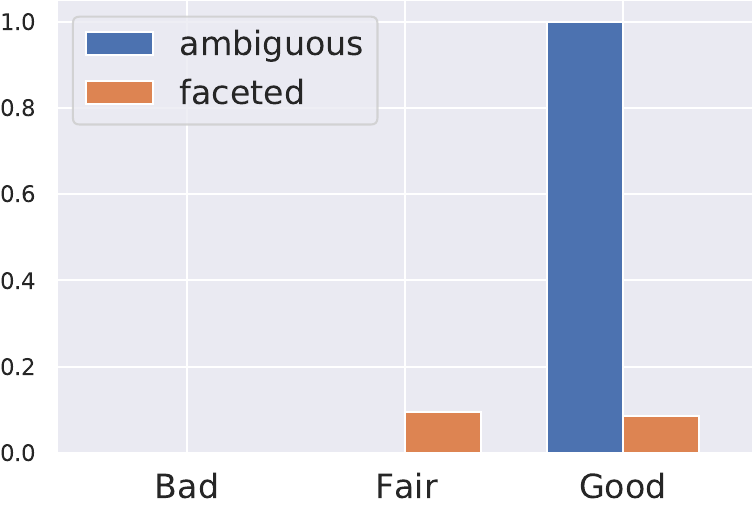} }}%
    \quad
    \subfloat[\centering MIMICS-Duo]{{\includegraphics[width=.45\columnwidth]{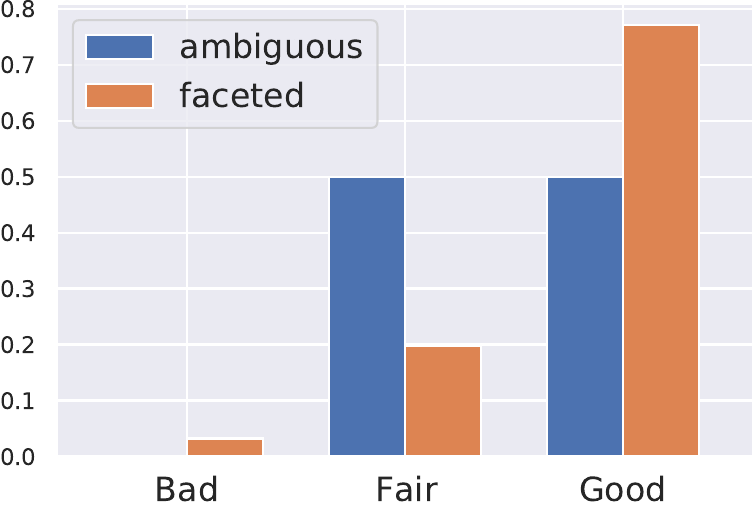} }}%
    \caption{clarifying question usefulness as per ambiguous or faceted queries.}%
    \label{fig:mimics_ambiguous_faceted}%
\end{figure}

\subsubsection{Subjectivity and Sentiment Polarity of CQs}
Next, we also argue that the subjectivity and sentiment polarity of a clarifying question can significantly impact its effectiveness. Subjectivity refers to the degree to which a question expresses a belief rather than objective facts. In the context of clarifying questions, highly subjective questions may provide the desired level of clarification since they reflect the perspective of the questioner and may resonate with the user's information needs.  Sentiment polarity, on the other hand, refers to the emotional tone of a question, typically measured as positive, negative, or neutral. In the context of clarifying questions, sentiment polarity can affect user satisfaction and engagement with the search system. Positive or neutral sentiment questions can make users feel more comfortable and encouraged to provide the needed information. However, negative sentiment questions may lead to user frustration or confusion, which can hinder the clarification process \cite{sekulic2021user}. In Figure~\ref{fig:features_correlation}, we include the correlation score between the calculated sentiment or subjectivity and the usefulness of the clarification. To calculate the sentiment and subjectivity, we use the TextBlob\footnote{\url{https://textblob.readthedocs.io/en/dev/}} package for Python which is a convenient way to do a lot of Natural Language Processing (NLP) tasks.

\subsection{Characterizing Queries with CQ Quality}

\subsubsection{Analyzing Clarification Quality upon Question \& Query Length}
The research literature suggests that longer queries often pose greater challenges in producing high-quality results \cite{zamani2020analyzing,aliannejadi2021analysing}. One reason for this is that longer queries may contain more irrelevant or ambiguous information, making it harder to match the user's intent with relevant results.

To answer \ref{RQ2}, which investigates the types of queries that require clarification, in Figure~\ref{fig:mimics_length_to_label}, we examine the clarification usefulness received by the clarification pane as a function of query and question length. 

Intriguingly, as the query length increases, there is a noticeable decline in the rate of clarification usefulness. In general, the results indicate that users are more satisfied with short queries and long clarifying questions, suggesting that shorter queries can potentially lead to more ambiguity, creating room for the system to intervene. In addition, the shorter queries increase the benefit of exploration and could further improve the level of user satisfaction with proper clarifying questions to retrieve the target information.

\begin{figure}%
    \centering
    \subfloat[\centering Features correlation with CQ]{{\includegraphics[width=.45\columnwidth]{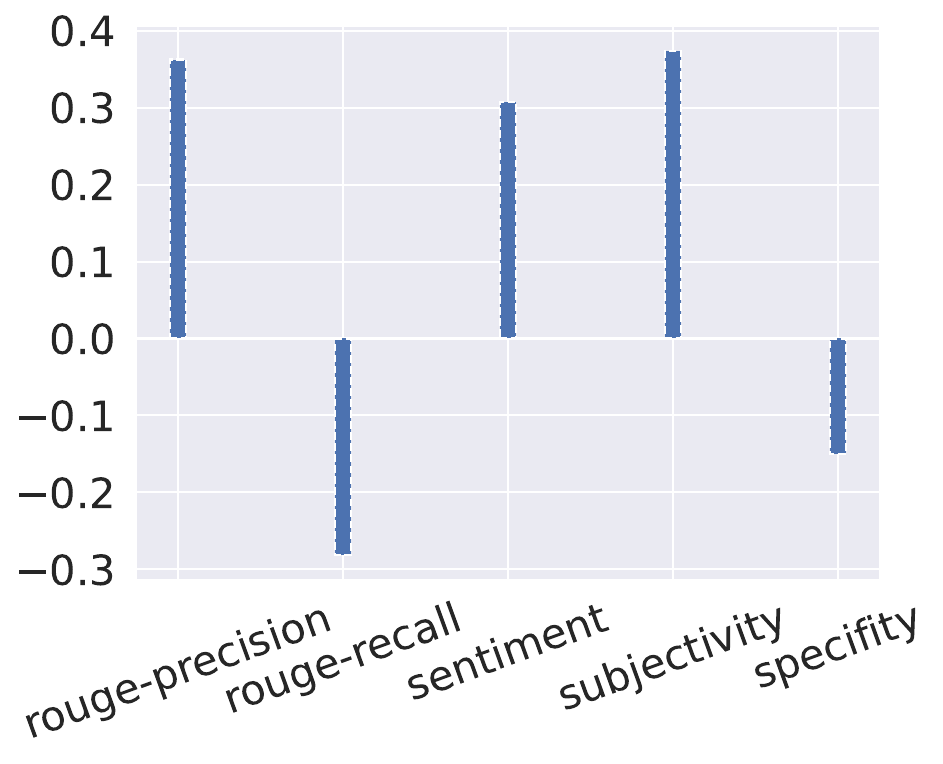} }\label{fig:features_correlation}}%
    \quad
    \subfloat[\centering User analysis]{{\includegraphics[width=.45\columnwidth]{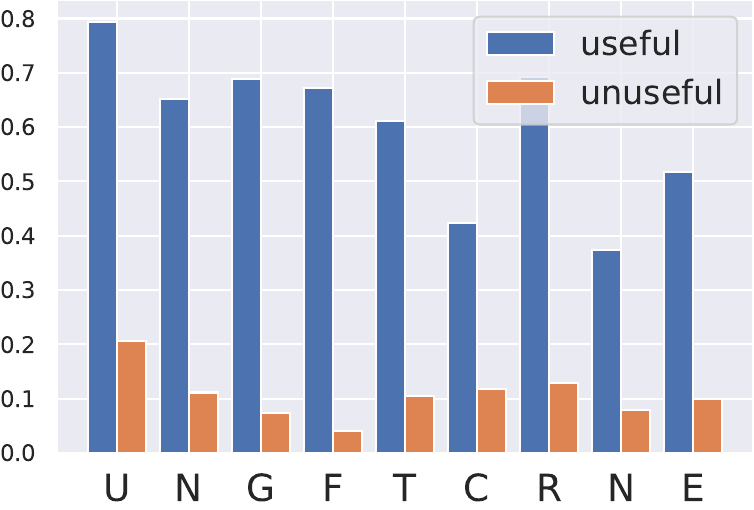}}}%
    \caption{Correlation evaluation of numerous features with clarifying question usefulness (left) and user study on if the usefulness of clarifying questions (right) can be determined by a given aspect. The aspects under evaluation include clarification-based aspects: `CQ Usefulness (U)', `Naturalness (N)', `Grammar correctness (G)', `Fluency (F)', `Template (T)', and joint modelling of query and CQs: `Coverage (C)', `Relevance (R)', `Novelty (N)', `Efficiency (E)'}%
    \label{fig:user_study}%
\end{figure}

\subsubsection{Ambiguous vs.~Faceted Queries}
In web search, clarifying questions can be valuable in uncovering the user's information needs behind ambiguous or faceted queries. To further answer \ref{RQ2}, Figure \ref{fig:mimics_ambiguous_faceted} illustrates the clarification usefulness rate for ambiguous and faceted queries. We define each query's category automatically based on the clarifying question templates and the candidate answers generated in the clarification pane. Ambiguous queries are those with multiple distinct interpretations, while facets are used to address underspecified queries by covering different aspects through subtopics \cite{AliannejadiSigir19,clarke2009overview}. According to the figure on MIMICS, clarifying questions for faceted queries are found to be more useful than those for ambiguous queries. However, on MIMICS-Duo, although faceted queries have a better rate, ambiguous queries also receive a remarkable usefulness rate. This suggests that for ambiguous queries, one query intent is more likely to dominate the user's information needs for the query --- usually the most popular one~\cite{DBLP:conf/emnlp/ProvatorovaBVK21}.

\subsubsection{Relevance Between Query and Questions} 
When measuring the usefulness of a clarifying question, it is intuitive that a clarifying question is required to be relevant to a given query. To reveal the correlation between such relevance and the usefulness of a clarifying question, we leverage the commonly used lexical-wise metric, Rouge scores, for analyzing such a feature. In Figure \ref{fig:features_correlation}, we present a correlation test result when using rouge-precision and rouge-recall. Rouge-recall refers to the proportion of important information that is captured by the generated clarifying questions, while rouge-precision refers to the proportion of generated questions that are relevant and useful in clarifying the user's request. Ideally, generated clarifying questions should have high recall (i.e., capture as much important information as possible) and high precision (i.e., only ask relevant and useful questions). We observe a noticeable positive impact of query-question relevance on the clarification usefulness while using the rouge-precision scores. Meanwhile, we also observe a negative correlation between the rouge recall scores and the clarification usefulness. These observations show that a clarifying question can be useful while capturing specific aspects of a given query. However, when the number of aspects covered within a clarifying question increases, the clarifying question becomes less useful (as per the negative correlated rough recall), which shows the negative impact of using general clarifying questions. These observations align with our findings in Section~\ref{ssec:template} about the usefulness of specific questions but general ones.

%% file: sections/04-results.tex
\section{Clarifying Question Usefulness Prediction}
\begin{table*}
    \centering
    \begin{adjustbox}{width=365pt}
    \begin{tabular}{lcccclcccl}
    \toprule
    \multirow{2}{*}{\textbf{Model}} & \multirow{2}{*}{\textbf{Type}} & \multicolumn{3}{c}{\textbf{MIMICS}} & \multirow{2}{*}{\textbf{impr.}} & \multicolumn{3}{c}{\textbf{MIMICS-Duo}} & \multirow{2}{*}{\textbf{impr.}} \\
    \cline{3-5}\cline{7-9}
     && Precision & Recall & F1 & & Precision & Recall & F1 & \\
     \midrule
     \multicolumn{10}{c}{Traditional Approaches} \\
     \midrule
     \multirow{2}{*}{RFC} & org. & 0.7522 & 0.5172 & 0.6129 & & 0.1256 & 0.2500 & 0.1672 \\
      & enr.~& \textbf{0.9474} & \textbf{0.9167} & \textbf{0.9318} & 52.2\%& \textbf{0.2560} & \textbf{0.3333} & \textbf{0.2896} & 73.2\% \\ \hline
     \multirow{2}{*}{DTC} & org. & 0.5648 & 0.5168 & 0.5397 & & 0.2218 & 0.2311 & 0.2263\\
      & enr. & \textbf{0.9288} & \textbf{0.9124} & \textbf{0.9205} & 70.6\% & \textbf{0.3291} & \textbf{0.3369} & \textbf{0.3330} & 47.1\%\\ \hline
     \multirow{2}{*}{SVC} & org. & 0.7360 & 0.5947 & 0.6578 && 0.2379 & 0.2498 & 0.2437\\
      & enr. & \textbf{0.8854} & \textbf{0.8830} & \textbf{0.8842} & 34.4\%& \textbf{0.3181} & \textbf{0.3321} & \textbf{0.3250} & 33.4\%\\
     \midrule
     \multicolumn{10}{c}{Neural Approaches} \\
     \midrule
     \multirow{2}{*}{BART} & org. & 0.9385 & 0.9310 & 0.9347 & & 0.3802 & 0.3762 & 0.3782 & \\
     & enr. & \textbf{0.9533} & 0.9271 & \textbf{0.9400} & 0.6\% & \textbf{0.9674} & \textbf{0.9186} & \textbf{0.9424} & 149.2\% \\ \hline
     \multirow{2}{*}{DBT}  & org. & 0.9348 & 0.9309 & 0.9328 & & 0.3709 & 0.3612 & 0.3660 & \\
       & enr. & \textbf{0.9473} & 0.9301 & \textbf{0.9386} & 0.6\% & \textbf{0.9698} & \textbf{0.9186} & \textbf{0.9435} & 157.8\% \\ \hline
     \multirow{2}{*}{BERT}  & org. & 0.9385 & 0.9310 & 0.9347 & & 0.3696 & 0.3721 & 0.3708 & \\
       & enr. & \textbf{0.9658} & \textbf{0.9479} & \textbf{0.9568} & 2.4\% & \textbf{0.9710} & \textbf{0.7441} & \textbf{0.8425} & 127.2\% \\
    \midrule
     \multicolumn{10}{c}{LLMs} \\
     \midrule
     \multirow{2}{*}{GPT-4} & org. & 0.3577 & 0.2149 & 0.2685 && 0.3061 & 0.2984 & 0.3022 \\
     & enr. & \textbf{0.3952} & \textbf{0.2839} & \textbf{0.3303} & 23.0\% & \textbf{0.3354} & \textbf{0.3228} & \textbf{0.3290} & 8.9\% \\ \hline
     \bottomrule
    \end{tabular}
    \end{adjustbox}
    \caption{The performance on user satisfaction prediction with CQs on MIMICS and MIMICS-Duo. RFC, DTC, DBT refer to the random forest, decision tree, and DistilBERT-based classifiers. The best models are in \textbf{bold}. `org.' and `enr.' indicate the basic implementation and feature-enriched implementation of approaches.}
    \label{tbl:user_satisfaction_results}
\end{table*}

\label{sec:results}

After exploring the correlation between available features and the usefulness of clarifying questions (CQs), in this section, we aim to answer \ref{RQ3} by evaluating the effectiveness of various features for predicting CQ usefulness. We consider both traditional ML and recent neural approaches discussed in Section~\ref{sec:experimental_setup} for the task of CQ usefulness classification, using query-question-candidate answer triplets as input on the MIMICS and MIMICS-Duo datasets.

To demonstrate the effectiveness of including additional CQ features for CQ usefulness prediction, we concatenate observed related features from Section~\ref{sec:analysis}, including CQ length, rouge-precision, sentiment polarity, and subjectivity, which are positively correlated with the clarifying question usefulness, with the original input for comparison. For the use of GPT-4 model, we carefully crafted a prompt to ask the model to generate a label-only output (good, fair or bad) with the query and clarifying question as input or with the inclusion of additional features. The corresponding prompt is provided in Appendix~\ref{sec:prompts}.

We present the experimental results in Table~\ref{tbl:user_satisfaction_results}. We observe that across the two datasets, incorporating our proposed features leads to large improvements on the traditional, neural approaches and large language models on both MIMICS and MIMICS-Duo datasets. In particular, the improvements to the traditional classifiers are significant, especially on the MIMICS dataset, with a minimum of 69.6\% and up to 151.4\% increases in F1 score. The resulting performance can also be comparable with advanced neural models. On the other hand, on the MIMICS-Duo dataset, by comparing the performance of the traditional classifiers with and without additional features as well as the basic neural models, their classification performances are less promising, which equally gives lower than 40\% of F1 scores (even the additional features can improve the basic traditional approaches with a minimum 45\% increase of F1 scores). However, by incorporating the positively correlated features into the neural model, we observe a significant impact (minimum 120\% improvement) on the model's performance, resulting in nearly perfect classification accuracy.  Meanwhile, as for the performance of the GPT-4 model, we observe that it does not perform competitively with the other two groups of approaches. The low accuracy of the GPT-4 model can be caused by its autoregressive nature of label generation, which does not guarantee a good classification outcome without fine-tuning. However, the use of additional features can still contribute to an improved performance of GPT-4, which further validates the effectiveness of using these positively correlated features.

\section{User Study Evaluation}
After observing promising performance improvements by including clarifying question features for usefulness estimation, we further conduct a user study to examine user opinions towards potential usefulness features by leveraging the expertise of domain experts in identifying potential relevant features for usefulness prediction. We identify eight additional features that can potentially advance usefulness prediction, divided into two groups: clarification features (i.e., naturalness, grammar, and fluency) that evaluate the text quality of a clarifying question and query-question features (i.e., coverage, novelty, efficiency, relevance, and question template) that measure if a clarifying question can effectively aid a query by addressing missing aspects, identifying novel but useful aspects, retrieving relevant documents or using particular templates.

We present 50 sampled query-clarifying question-feature triplets to seven domain experts to annotate the usefulness of CQs. We then ask them to label which features are most essential for considering a CQ useful. Also, we ask them to select the minimum-required features for a CQ to be deemed useful. We summarize and present the user study results in Figure~\ref{fig:user_study} (b).
The results of the study show that a high textual quality question is necessary for a CQ to be considered useful, especially in terms of naturalness. Additionally, among the query-question features, relevance is commonly considered an issue that needs to be addressed to present useful CQ. This observation aligns with previous efforts in the literature that link query aspects with CQs to generate them effectively \cite{zamani2020generating}. Another interesting finding is that coverage is one of the lowest-scored features, which also aligns with our previous consistent findings on using specific, rather than high aspect-recall clarifying questions. Therefore, we conclude that the user study further highlights the value of the text quality of CQs and their relevance to queries, in addition to the features such as length, subjectivity and specificness that we previously identified as useful through experimental results. 

\label{sec:exp_satisfaction_appendix}

%% file: sections/05-conclusion.tex
\section{Conclusion and Future Work}
\label{sec:conclusion}

This paper analyzed the usefulness of clarifying questions using two well-known real-world clarifying question datasets.
Specifically, we studied the impact of various features related to both clarification questions and the corresponding query on the usefulness of clarifying questions with respect to the level of user satisfaction. The analytical results indicate the positive impact of having specific, positively sentimental-oriented, lengthy and subjective clarification questions. By leveraging such analysis, we introduce these positively correlated features to the usefulness estimation of clarification questions. As per the classification accuracy, we observed a consistent improvement in applying the additional features, especially on the traditional approaches, with a minimum 45.7\% improvement. Furthermore, the performance-boosting on the neural approaches enables the classifiers to achieve a consistent, nearly perfect performance with over 94\% classification precision.

The results of our usefulness prediction models proved our hypothesis that incorporating different feature types would help improve the prediction by a large margin. In addition, we augment our contributions with another user study, which uses users' opinions in examining the usefulness of clarification questions from various perspectives, and we also observed close conclusions with our experimental findings.

In the future, we plan to further study the impact of the features on pre-trained language models and explore various methods such as prompting large language models to generate useful and satisfying clarifying questions. Furthermore, we plan to fine-tune open-source large language models such as LLaMA \cite{touvron2023llama} to select the more relevant and useful clarifying questions between several questions when a model generates more than one clarifying question to clarify users' ambiguity.